\documentclass{article}
\usepackage{chicago}
\usepackage{amssymb}

\newcommand{\ie}{\mbox{i.\,e.\,\ }}
\newcommand{\iec}{\mbox{i.\,e.\,}}

\newcommand{\egc}{\mbox{e.\,g.\,}}

\newcommand{\vctr}[1]{\ensuremath{\mathbf{ #1 }}}

\newcommand{\dr}[1]{\ensuremath{\mathrm{d} #1\,}}
\newcommand{\mc}[1]{\ensuremath{\mathcal{#1}}}

\newcommand{\dbd}[2]{\ensuremath{\frac{\dr{#1}}{\dr{#2}}}}

\newcommand{\pbp}[2]{\ensuremath{\frac{\partial #1}{\partial #2}}}

\newcommand{\bk}[2]{\ensuremath{\left\langle #1 | #2 \right\rangle}}

\newcommand{\op}[1]{\ensuremath{\widehat{\textsf{\ensuremath{#1}}}}}

\newcommand{\be}{\begin{equation}}
\newcommand{\ee}{\end{equation}}
\newcommand{\e}[1]{\mathrm{e}^{#1}}

\newcommand{\re}{\ensuremath{\mathbf{R}}}
\newcommand{\AdS}{\mathrm{AdS}}
\newcommand{\CFT}{\mathrm{CFT}}

\begin{document}
\title{The case for black hole thermodynamics \\  Part II: statistical mechanics}
\author{David Wallace\thanks{Dornsife College of Letters, Arts and Sciences, University of Southern California; email \texttt{dmwallac@usc.edu}}}
\maketitle

\begin{abstract}
I present in detail the case for regarding black hole thermodynamics as having a statistical-mechanical explanation in exact parallel with the statistical-mechanical explanation believed to underly the thermodynamics of other systems. (Here I presume that black holes are indeed thermodynamic systems in the fullest sense; I review the evidence for \emph{that} conclusion in the prequel to this paper.) I focus on three lines of argument: (i) zero-loop and one-loop calculations in quantum general relativity understood as a quantum field theory, using the path-integral formalism; (ii) calculations in string theory of the leading-order terms, higher-derivative corrections, and quantum corrections, in the black hole entropy formula for extremal and near-extremal black holes; (iii) recovery of the qualitative and (in some cases) quantitative structure of black hole statistical mechanics via the AdS/CFT correspondence. In each case I briefly review the content of, and arguments for, the form of quantum gravity being used (effective field theory; string theory; AdS/CFT) at a (relatively) introductory level: the paper is aimed at readers with some familiarity with thermodynamics, quantum mechanics and general relativity but does not presume advanced knowledge of quantum gravity. My conclusion is that the evidence for black hole statistical mechanics is as solid as we could reasonably expect it to be in the absence of a directly-empirically-verified theory of quantum gravity.
\end{abstract}

\section{Introduction}\label{introduction}

In the last forty years, the evidence has become overwhelming that stationary or near-stationary black holes behave, as far as any description by external observers is concerned, exactly like ordinary thermodynamic systems. (I review this evidence in depth in \citeN{wallaceblackholethermodynamics}, the prequel to this paper --- henceforth `Part I'). Right from the outset --- and before the discovery of Hawking radiation~\cite{hawking1975}, and the development of the membrane paradigm~\cite{membraneparadigm}, really solidified the case for black hole thermodynamics --- it was conjectured that black hole thermodynamics has the same ultimate justification as the rest of thermodynamics: in statistical mechanics. In particular, if (in Planck units) one-quarter the surface area $\mc{A}$ of a black hole really can be identified with entropy, there ought to be a space of microstates of that black hole, all semiclassically indistinguishable, of dimension $\exp(\mc{A}/4)$.

In the view of (most of) the theoretical physics community, this is no longer a conjecture. Instead (it is widely said) a great deal of evidence has now been collected in support of a black hole statistical mechanics (BHSM) underpinning black hole thermodynamics. This evidence is of course not empirical (there is no empirical evidence as yet for Hawking radiation, let alone for BHSM); rather, it consists of reproductions of the semiclassical formulae for black hole entropy in those partly-developed theories of quantum gravity which we currently possess, to a precision that would be difficult or impossible to make sense of unless those `partly-developed theories' really were partial descriptions of a consistent quantum theory of gravity that sufficed to ground black hole thermodynamics.

In this paper, I want to review this evidence, which to my eyes is extremely convincing, for the benefit of the somewhat-sceptical non-specialist. Other reviews exist, and indeed I drew heavily on them in writing this paper (see in particular Carlip~\citeyear{carlipreview,carlipreview2}, \citeN{harlowreview}, \citeN{hartmanreview},  \citeN{polchinskiblackholereview}) but, as with the case for black hole \emph{thermodynamics}, they tend to be focussed on recent results and to assume that the reader is antecedently disposed to accept BHSM at face value. I confine my attention to physics outside the black hole event horizon, and do not directly consider the so-called `information-loss paradox', although (as I argue in more detail in \citeN{wallaceinformationloss}) it is ultimately the validity of a statistical-mechanical account of black holes that would make information loss so paradoxical. For similar reasons, I do not discuss `horizon complementarity' or the `firewall paradox'; again, see \citeN{wallaceinformationloss} and references therein for more on these topics.

The structure of the paper is as follows. In section \ref{generalprogram} I review statistical mechanics and then consider what the general structure of a statistical-mechanical theory of black holes would look like. In sections \ref{lowenergy}--\ref{adscft} I successively review the evidence for BHSM coming from, respectively, general relativity as an effective field theory; string theory; the AdS/CFT correspondence. Section \ref{conclusion} is the conclusion.

As with Part I, I adopt a level of mathematical rigor roughly the same as that in the mainstream theoretical physics literature, and so fall short of the fully-rigorous methods often found in foundational discussions. This seems to me basically unavoidable for BHSM: the theories we are studying are not yet fully articulated and so complete rigor is premature. Again as with Part I, I hope that readers dissatisfied with this will at least take away some understanding of why physicists themselves take BHSM so seriously. 

I adopt units where $G=k=c=\hbar=1$ unless explicitly stated otherwise. $l_p=\sqrt{\hbar G/c^3}$ denotes the Planck length; note that in units where $\hbar=c=1$, $G=l_p^2$.

\section{Black hole statistical mechanics: the general program}\label{generalprogram}

Thermodynamics, in of itself, is a purely phenomenal theory: its postulates are simply assumed, as are the equations of state for any given system. Statistical mechanics aims to provide a microphysical underpinning for thermodynamics, justifying its foundational postulates and permitting derivations of its thermodynamic features from the microphysics of the system being studied. (This does not exhaust its content, which also includes the calculation of  non-thermodynamic features of equilibrium and near-equilibrium states such as correlation lengths and transport coefficients, the quantitative features of the approach to equilibrium, and the analysis of complex systems in contexts where `equilibrium' is not a helpful concept.) To get clear on what it means for black holes in particular to have a statistical-mechanical description, it will be helpful to start by reviewing how such descriptions work in general. (This is standard physics and I do not provide explicit references; see any textbook account of statistical mechanics.)

\subsection{A brief review of statistical mechanics}

Much of the foundations of statistical mechanics is contested, but at least the following are clearly parts of statistical mechanics as used in calculational physics. Firstly, the various irreversible features of thermodynamics are consequences of unitary (or, for classical systems, Hamiltonian) dynamics, together with some (often-tacit) assumptions about a system's initial conditions and possibly some probabilistic assumptions. In particular, the approach of a system to equilibrium is studied quantitatively in non-equilibrium statistical mechanics through equations like the Boltzmann and Langevin equations, which are constructed from the microphysics.

Secondly, the equilibrium equation of state is derived within statistical mechanics via a microphysical expression for the entropy, which makes the latter a measure of the Hilbert-space dimension associated with an equilibrium system. This can be done in two main ways: via the \emph{microcanonical ensemble}, where an equilibrium system is treated as being genuinely isolated from its environment and its thermodynamic energy is identified as its microphysical entropy, or via the \emph{canonical ensemble}, where the system is treated as being in thermal contact with a very large reservoir at a fixed temperature and the system's thermodynamic entropy is a probabilistic average of its microphysical energies. 

In more detail: given a quantum system with Hamiltonian \op{H} and, say, conserved charge $\op{Q}$ and angular momentum $\op{J}$, we can define a subspace $\mc{S}(\Delta_H,\Delta_Q,\Delta_J)$ of all joint eigenstates of these operators with eigenvalues for  $\op{H}$, $\op{Q}$ and $\op{J}$ lying in $\Delta_H$, $\Delta Q$ and $\Delta_J$ respectively. Then the \emph{density of states} $N(E,Q,J)$ is defined by the expression
\be
N(E,Q,J)\delta E \delta Q\delta J = \mathrm{Dim} \,\,\mc{S}([E,E+\delta E],[Q,Q+\delta Q],[J,J+\delta J])
\ee
where the quantities $\delta E$, $\delta Q$, $\delta J$ are large enough to include many eigenstates but small compared with macroscopically relevant scales. In the microcanonical approach, the entropy is defined as the logarithm of the density of states, 
\be
S(E,Q,J)= \ln N(E,Q,J)
\ee
and the thermodynamic temperature is 
\be
T = \left(\pbp{E}{S}\right)_{Q,J}
\ee
while in the canonical approach, the  most convenient starting point is the \emph{partition function}
 \be \label{partitionfunction}
Z (\beta,Q,J) = \int \dr{E}\e{-\beta E} N(E,Q,J) \equiv \int \dr{E} \e{S(E)-\beta E}
 \ee
 where $\beta=1/T$ is the inverse thermodynamic temperature, and the expected energy and entropy can be calculated from the partition function by
 \be
 \langle E \rangle = -\dbd{}{\beta}(\ln Z);\,\,\,\,\, S_C = \beta \langle E \rangle + \ln Z
 \ee 
 (where I write the canonical entropy as $S_C$ to distinguish it from the microcanonical entropy).

Under certain conditions, these two ensembles give the same results as one another for the values of thermodynamic quantities like energy and entropy; partly because the calculational methods will recur later, I briefly review those conditions. In general, if we have an integral
\be
I = \int \dr{x} \e{-f(x)}
\ee
where $f$ has a minimum at $x_0$ and increases steeply away from that minimum, we can approximate the integral as
\be
I \simeq \e{-f(x_0)}\int \dr{x}\e{-f''(x_0)(x-x_0)^2/2} = \frac{\e{-f(x_0)}}{\sqrt{\pi f''(x_0)}}
\ee
(this is the \emph{saddle-point method} for evaluating such an integral). Applying this method to the function $f(E)=\beta E- S(E)$, and assuming it has its minimum for fixed $\beta$ at $E=E_0$ we can evaluate
\be
Z(\beta) \simeq \frac{\e{S(E_0)-\beta E_0}}{\sqrt{\pi S''(E_0)}}
\ee
and, by the same method,
\be
\langle E \rangle \simeq E_0;\,\,\,\, \mbox{Var}(E) \simeq 1/S''(E_0)
\ee
under the assumption that 
$ \mbox{Var}(E)\ll \langle E \rangle ^2.$ This can be understood physically by the (exact) relation
\be \label{heatcapacity}
C_V = \pbp{\langle E \rangle}{T} = \frac{\mbox{Var}(E)}{T^2}
\ee
where $C_V$ is the (canonical) heat capacity at constant external parameters. So the condition for the validity of these approximations is 
\be
\label{limitcondition} (T/\langle E \rangle)^2 C_V \ll 1
\ee
 and under that approximation we have
\be
\ln Z(\beta) \simeq S(\langle E \rangle) - \beta \langle E \rangle - \frac{1}{2}\ln (\pi C_V)
\ee
from which follows $S_C(\beta) \simeq S(\langle E \rangle)+$ constant. So when the condition (\ref{limitcondition}) is met, we can work exchangeably with the canonical or microcanonical ensembles. In this situation, the system is said to be at the \emph{thermodynamic limit}. (This must be distinguished from other, related uses of the term, such as to refer to the infinite-volume limit.)

\subsection{Statistical mechanics of black holes}

As I discussed in Part I, self-gravitating systems have unusual thermodynamic properties: in particular, the long-range forces that bind them tend to lead to negative heat capacities. No such system can be in stable equilibrium with a heat reservoir: an arbitrarily small fluctuation of heat from the reservoir to the system will cool the system and so lead to further heat transfer, so that heat will flow into the system without limit; an arbitrarily small initial fluctuation in the opposite direction will have the opposite instability, with the system heating up as it emits more and more heat. So this suggests that a strongly self-gravitating system at equilibrium cannot normally be analysed by the canonical distribution. And this follows formally from (\ref{heatcapacity}), which demonstrates that the canonical heat capacity is always positive.

But the \emph{micro}canonical distribution is in principle perfectly well defined for a self-gravitating system in isolation --- say, in a very large box with reflecting walls. At equilibrium, that box will be filled with thermal radiation, and we can think of the system as consisting of the self-gravitating system without radiation, in equilibrium with that radiation. That in turn requires our box not to be \emph{too} large, else it will again be unstable; as we saw in Part I, for a black hole in particular the mass of the radiation must be less than 1/4 the mass of the black hole, which is compatible with extremely large boxes for astrophysical-scale black holes.

As observed by \citeN{yorkcanonical}, though, we can still use canonical-ensemble methods to discuss a black hole if we place it in a sufficiently \emph{small} box, say a spherical box of radius $r$. The reason (specialising for convenience to a Schwarzschild black hole) is that the local temperature of the black hole's Hawking radiation at distance $r$ is given by $T(r)=T_H(M)/\alpha(M,r)$, where $T_H(M)=1/8\pi M$ is the Hawking temperature at mass $M$, and $\alpha(M,r)=(1-2M/r)^{-1/2}$ is the redshift at distance $r$. The boundary of the box will be at equilibrium with the hole if it is at temperature $T(r)$. Adding mass to the hole decreases both $T_H(M)$ and $\alpha(M,r)$, and for a sufficiently small box the latter effect can dominate. Indeed, an elementary calculation gives
\be
\pbp{M}{T}=8\pi M^2 \left(\frac{r-2M}{3M-r}\right)
\ee
which is positive when $r$ lies outside the Schwarzschild radius $r=2M$ but inside the surface of last stable orbit $r=3M$. So in this regime we can use the canonical ensemble to describe the black hole. In the same regime 
\be
(T/M)^2\pbp{M}{T} = \left(\frac{1}{8 \pi M}\right)^2  \left(\frac{r-2M}{3M-r}\right)
\ee
which will be $\ll 1$ for $M \gg m_p$, so the conditions for equivalence of canonical and microcanonical ensembles will be satisfied unless the box radius is extremely close to $3M$. (A slight subtlety is that for a box of finite size, the energy --- defined in terms of the boundary Hamiltonian on the box --- deviates somewhat from $M$, but not enough to invalidate this analysis. See York \emph{ibid} for details.)

Whichever method we use to define BHSM, the basic premise will be that a black hole of mass $M$, charge $Q$ and angular momentum $J$ is (or at any rate can be understood as) a quantum-mechanical system with density of states
\be \label{densityofstates}
N(M,Q,J) = K \exp S_{TD}(M,Q,J)
\ee
where $S_{TD}$ is the thermodynamic entropy of the black hole as calculated within black hole thermodynamics (see Part I) and $K$ is some constant such that $\ln K$ is small compared to the area of a classical-scale black hole. If we assume the Einstein field equations, $S_{TD}$ is one-quarter the area of the event horizon; for a Schwarzschild black hole in particular, we have
\be
N(M,0,0) \propto\exp (4 \pi M^2).\ee
(Note that for a large box, where $M$ can be treated as the energy, the partition function clearly diverges for this density of states, validating our previous arguments.)

If so, a black hole would have a very large number of microscopic degrees of freedom, none of which show up in the semiclassical description, and the obvious question is: where are they located? The area formula for black hole entropy is suggestive: if black hole entropy is proportional to area, it suggests putting quantum degrees of freedom, one per every few Planck areas, on or near the horizon (but outside it, else they could play no explanatory role in a black hole's thermodynamic behaviour). But we can do much better than this simple heuristic. Specifically, we saw in Part I (following \citeNP{membraneparadigm}) that 
the thermodynamics of black holes can be elegantly expressed via the
\begin{description}
\item[Membrane paradigm:] With respect to any semiclassical or thermodynamical physical processes taking place outside the stretched horizon of a stationary or near-stationary black hole, that horizon may be treated as a thermodynamical system (specifically, as a thin membrane of conducting fluid) at or near thermal equilibrium.
\end{description}
(Recall that the stretched horizon is a timelike surface approximately one Planck length from the true horizon.) In particular, the stretched horizon is in thermal equilibrium with the black hole atmosphere (the layer of Hawking radiation largely trapped behind the angular-momentum potential barrier), at a locally-measured temperature of order the Planck temperature. So to provide a statistical-mechanical underpinning to BHT, we need to posit the
\begin{description}
\item[Quantum membrane paradigm (QMP):] With respect to \emph{any} physical process taking place outside the stretched horizon of a stationary or near-stationary black hole, that horizon may be treated as a quantum-mechanical system at or near thermal equilibrium (as defined by the asymptotic Hamiltonian), with density of states given approximately by $N \propto \exp(S_{TD})$.
\end{description}
(The reason for the ``approximately'' in this description is that some fraction of the black hole's entropy will be comprised of the atmosphere's entropy; I return to this point in section \ref{oneloop}.) The quantum membrane paradigm was originally proposed by  \citeN{susskindthorlaciusuglum} following related proposals by \citeN{thooftblackhole}, though the name is mine (`complementarity' is the most common term used in the literature, but I avoid that term as it has other connotations). 

\section{Low-energy quantum gravity}\label{lowenergy}

Already there is an interesting tension visible in BHSM. Reproducing the Bekenstein-Hawking entropy requires a density of states so high as to imply that Planck-scale physics is needed to understand those states. Yet the entropy formula itself is derived within ordinary general relativity applied at quite large scales and it is hard to see how its numerical value could be sensitive to the smallest-scale details of the quantum theory of gravity to which general relativity is presumably an approximation.

We can get insight into this puzzle, as well as our first direct evidence for BHSM, by considering \emph{quantum} general relativity in the low-energy regime, \ie where the metric is treated quantum-mechanically but curvatures remain low compared to Planck scales. Quantum gravity is reasonably well understood in this regime via the path-integral formulation of quantum field theory and the effective-field-theory method for understanding renormalisation; I begin by reviewing these techniques. The account is realistically too terse for readers completely unfamiliar with the subject but it should serve as a reminder to readers with some previous acquaintance, and also to bridge the gap between QFT as used in black hole thermodynamics and as seen in other areas of physics; again, it is established physics and I omit original references. (For the full picture, see, \egc, \citeN{banksQFT}, \citeN{duncanQFT}, \citeN{srednickiqft}, or \citeN{zeeQFT}; I give a more in-depth review of the concepts aimed at non-specialists in \citeN{wallaceqfthandbook}.)

\subsection{Quantum field theory: a brief review}\label{qftintro}

The central claims of QFT, in the `effective field theory' form relevant for our purposes, are:
\begin{enumerate}
\item The dynamics of the theory are expressed in terms of a collection of classical fields, and a classical action for those fields. Using a scalar field theory on flat background spacetime with metric $\eta$ as an example, a fairly general action (on some spacetime region $M$) for such a field is given by 
\be
S[\varphi] = \int_M \mc{L}(\varphi, \nabla \varphi) = \int_M \left(\frac{1}{2}\eta (\nabla \varphi,\nabla \varphi) - m^2 \varphi^2 - \sum_{n=2}^\infty \frac{\lambda_{2n}\varphi^{2n}}{(2n)!}. \right)
\ee
Time evolution is then given by the \emph{path integral}:
\be \label{pathintegral}
\bk{\phi_1,\Sigma_1}{\phi_2,\Sigma_2} = \int^{\phi_2}_{\phi_1} \mc{D}\varphi \exp (-i S[\varphi])
\ee
For the purposes of BHSM, it is more useful to analytically continue this expression to imaginary values of time to obtain an expression for the partition function:
\be
Z(\beta) = \int_\beta \mc{D}\varphi \exp (-S_E[\varphi])
\ee
where the $\beta$ subscript indicates that the path integral is over paths periodic in imaginary time with period $\beta$, and where $S_E$ is the \emph{Euclidean action}, obtained from $S$ by analytic continuation. For field theories with a gauge symmetry (in this context, this means: a transformation of the fields which leaves invariant both the boundary conditions and the action) these path integrals should in the first instance be taken only over equivalence classes of gauge-equivalent fields; it is normally possible to recover a path integral over all the fields, but often at the cost of introducing new auxiliary fields and/or modifying the action.
\item These path integrals are in general formally divergent, unless regularised by requiring the fields to vary on scales no smaller than some \emph{cutoff length} $\Lambda$. (This requirement can be imposed by, for instance, integrating over only Fourier modes with frequencies $<1/\Lambda$, or by putting the integral on a lattice of spacing $\Lambda$.) This cutoff can be interpreted \emph{physically} as a limit to the domain in which the field theory is reliable; the term \emph{effective field theory}, often used to describe QFTs in modern physics, is intended to convey that the theory is supposed to apply in certain regimes only and is not a candidate for `fundamental' physics. 
\item The effect of changing the cutoff from $\Lambda$ to some $\Lambda'$ is to rescale the infinite number of coefficients that specify the theory's Lagrangian ($m^2$ and $\lambda_4, \lambda_6, \ldots,$ for my scalar-field example). This transformation is known as the \emph{renormalisation group}. In general, renormalisation group flow affects the values of all coefficients in the action, even those initially equal to zero; for this reason, it is strictly speaking necessary to work with the most general Lagrangian compatible with a theory's symmetries. 
\item Some, but not all, field theories have non-trivial \emph{fixed points} in the space of coefficients, to which the coefficients tend under renormalisation-group flow in the limit of longer and longer cutoffs. If a theory has a nontrivial fixed point, the precise value of the cutoff is irrelevant for physics on scales long compared to the cutoff, and we can \emph{formally} understand the theory as defined on all lengthscales. In the absence of a fixed point (or if the only fixed point corresponds to an interaction-free theory) then the cutoff is directly relevant to the phenomenology. 
\item The coefficients can be divided into two categories on dimensional-analysis grounds: a finite collection of \emph{renormalisable} coefficients, and an infinite number of \emph{non-renormalisable} coefficients. For physical processes on a scale $L \gg \Lambda$, the effects of the nonrenormalisable coefficients (at least as calculated in perturbation theory) is suppressed by some positive power of $(\Lambda/L)$. As such, on such scales the nonrenormalisable coefficients can typically be disregarded; physically significant results are expressible in terms of the renormalisable coefficients.
To a first approximation, a QFT has a nontrivial fixed point iff it has renormalisable coefficients corresponding to interactions. To a somewhat better approximation, having such coefficients is at least a necessary condition for a fixed point: both scalar field theory and quantum electrodynamics are suspected of having no non-trivial fixed points even though they are renormalisable in the power-counting sense. Even this better approximation is not a rigorous result: the existence of delicate cancellations in the renormalisation process could in theory lead to a fixed point even in this case (this is known as \emph{asymptotic safety}~\cite{weinbergasymptotic} though there are at present no uncontentious empirically-relevant examples).
\item Practical calculations in QFT normally involve perturbation theory, where the solutions are found from perturbation around exactly-solvable cases. In the case of scalar field theory, for instance, if we restrict attention to renormalisable terms then the theory is specified by a mass $m$ and an interaction strength $\lambda_4$. The case $\lambda_4=0$ corresponds to a free theory, for which the path integral can be evaluated exactly. We might then expect to be able to express the path integral as a power series in $\lambda_4$, consisting of successive correction terms to the free-theory predictions. If this is attempted, though, the ``successive correction terms'' turn out to be very large, invalidating the assumptions of perturbation theory. Instead, the power series has to be expressed in terms of the \emph{renormalised} coefficients $m^{ren}$, $\lambda_4^{ren}$, which include finite (and cutoff-dependent) corrections. (Indeed, it is these renormalised coefficients which are measured empirically; the `bare' coefficients in the original Lagrangian are not empirically accessible.)
\item We can approximate the path integral for $Z(\beta)$, in particular, by saddle-point methods: if the Euclidean action has a single minimum at $\varphi=\varphi_0$, we can write $\varphi = \varphi_0 + \delta \varphi$ and expand it as
\be
S[\varphi]=S[\varphi_0] + S_2[\delta \varphi] + \mbox{higher terms}
\ee
to obtain
\be \label{pathint-approx}
\ln Z = S_E(\varphi_0) + \int_\beta \mc{D}\delta\varphi \exp(- S_2[\delta \varphi]) + \mbox{higher terms}.
\ee
(Essentially this method is also used in most other applications of perturbation theory to QFT.) 
In non-gravitational contexts, the term $S_E(\varphi_0)$ in (\ref{pathint-approx}) can be disregarded (it is normally zero and in any case can be absorbed into the normalisation of the path integral). The path integral is quadratic in the perturbation $\delta \varphi$ and can be interpreted as the contribution of a gas of free particles to the partition function. (For reasons originating in the Feynman-diagram techniques used to evaluate path integrals, these two stages of approximation are known as `zero-loop' and `one-loop' approximations to $Z$, respectively.) The higher-order terms represent successive corrections from interactions between the particles.
\end{enumerate}

\subsection{General relativity from an effective-field-theory viewpoint}

The classical action of vacuum general relativity (suppressing the cosmological constant for simplicity) is the \emph{Einstein-Hilbert action}:
\be
S_{E-H}[g]=\int_M \sqrt{-g} \frac{1}{8\pi G}  R 
\ee
where $R$ is the scalar curvature and I have restored Newton's constant $G$. (In this section I follow the normalisation conventions of \citeN{demersetal}.) Regarded as an action defining a QFT, this action is non-renormalisable, which at one point was the usual answer given as to why quantizing gravity by QFT methods was impossible. But from the effective-field-theory viewpoint there is nothing problematic about gravity being non-renormalisable; it simply points to\footnote{Points to, but does not entail, because of the possibility of asymptotic safety. The asymptotic-safety program in quantum gravity, which lies beyond the scope of this paper, aims to demonstrate the consistency of QFT versions of general relativity at arbitrarily high energies (see \citeN{reutersaueressig} for a review).} a breakdown of the theory at a lengthscale $\sim \sqrt{G}\equiv l_p$. (Indeed, the standard result that nonrenormalisable interactions are extremely weak at energies far below their cutoff can be seen as offering an explanation, of sorts, for the weakness of the gravitational force.) From this viewpoint (which is standard in most of theoretical physics, and which I adopt hereafter) the `problem' of quantum gravity arises not whenever we try to combine quantum mechanics and general relativity, but only when we try to do so in regimes where the energy scales approach Planck scales (\iec, in the early Universe and near the singularities within black holes --- and at the stretched horizon of a black hole, if we describe the physics from the perspective of a fiducial observer). Indeed, the low-energy version of quantum gravity obtained by treating general relativity as an effective field theory arguably even has (admittedly crude) observational support, since it is the theoretical basis for  calculations of the quantum fluctuations in the early Universe that are believed to be the origin of the fluctuations in the cosmic background radiation (see, \egc, \citeN[pp.470-474]{weinbergcosmology}). For a review of this effective-field-theory understanding of quantum gravity, see, \egc, \citeN{burgesseffectivegr}.

(At this point it will be helpful to establish some terminology.  By `low-energy quantum gravity' I mean general relativity, regarded as an effective field theory and applied only when the energy scales are far below the Planck scale. By `full quantum gravity' I mean some finite, cutoff-free theory valid even in Planck-scale regimes, which reduces to low-energy quantum gravity in appropriately limited regimes.)

The Einstein-Hilbert action alone is not quite suitable for constructing low-energy quantum gravity, for two reasons. Firstly, we have seen that QFT requires us to consider the most general action for a given field's symmetries, so we need to add to the scalar curvature term all other diffeomorphism-invariant terms expressible in terms of the metric. This leads to an action of form
\be \label{generalisedaction}
S[g] = \int_M \sqrt{-g} \left( \frac{1}{8\pi G}  R  + \frac{\alpha}{4\pi} R^2 + \frac{\theta}{4\pi}R_{ab}R^{ab} + \frac{\gamma}{4\pi}R_{abcd}R^{abcd}+\cdots \right).
\ee
Successive terms in this expansion have their effects on the physics at lengthscale $L$ suppressed by ever-higher powers of $\Lambda/L$. (In some versions of full quantum gravity, supersymmetry means that some of these terms vanish, but there will always be infinitely many left.)

Secondly, since the curvature depends on second derivatives of the metric, the classical equations of motion are obtained from variations that leave the metric \emph{and its normal derivative} invariant on the boundary, whereas the path integral boundary condition does not involve normal derivatives. This can be resolved by adding a boundary term to the action (the form of which is fixed by the `bulk' part of the action): in the case of the Einstein-Hilbert action, for instance, the boundary term is proportional to the extrinsic curvature of the metric at the boundary.

\subsection{Statistical mechanics of black holes}

With these modifications made, let's consider (in general relativity coupled to some matter field $\varphi$, and following \citeN{gibbonshawking} and \citeN{gibbonshawkingperry}) evaluating the partition function  of a system enclosed within a (spherically-symmetric) box of surface area $A_{box}$, at inverse temperature $\beta$. To do so by saddle-point methods (and no better method is currently available) we need to find the extrema of the action, and one obvious extremum is given by flat space, with matter fields vanishing and $g$ equal to the Minkowski metric. The value of the action at this extremum is zero (by construction: this is our convention for normalising the path integral), so the contribution to the partition function from this extremum, to one-loop order, has form
\be
\ln Z_{\mbox{flat}}(\beta) =  \int_\beta \mc{D}\delta g \mc{D}\varphi \exp(- S_{2,0}[\delta g,\varphi])
\ee
where $S_{2,0}$ is the second functional derivative of the action evaluated at empty flat space.
This has a fairly straightforward interpretation as the partition function of a gas of thermal particles: the quanta associated with the matter field, and the gravitons associated with metric fluctuations. 

(There is one technical subtlety: both in this case and generically, the extrema of the action are not \emph{minima} of the action because sufficiently rapid conformal fluctuations of the metric will decrease the action, which would seem to invalidate the use of saddle-point methods. The original applications of the path integral \cite{gibbonshawkingperry} dealt with this problem formally by rotating the contour of integration over those modes, and there is considerable evidence (\citeNP{hartleschleich}, \citeNP{schleich} that this rotation is justified by considerations of gauge-invariance, but the matter remains somewhat obscure.)

But this is not the \emph{only} extremum of the action. It is also extremised if the metric inside our surface is the (Euclidean analytic continuation of the) Schwarzschild black hole, with mass given by $M=\beta/8 \pi$. So there is an additional contribution $Z_H(\beta)$ to the partition function, given to one-loop order by
\be \label{blackholepartition}
\ln Z_H(\beta) = S_E(g_M,0) + \int_\beta \mc{D}\delta g \mc{D}\varphi \exp(- S_{2,M}[\delta g,\varphi])
\ee
where $g_M$ is the metric for the mass-$M$ Euclidean black hole, and $S_{2,M}$ is the second-order variation of the action around the extremum $(g=g_M,\varphi=0)$. 

The obvious temptation is to interpret this term as the contribution to the entropy due to the state space of a mass-$M$ black hole. But there are subtleties in doing so. In Gibbons and Hawking's original calculations the surface of the box was taken at lying at a radius far larger than the Schwarzschild radius 2$M$, and we have seen that the canonical ensemble is not well-defined in this situation. This shows up formally in a family of \emph{non}-conformal fluctuations that decrease the Euclidean action, again invalidating the use of the saddle-point method. \citeN{gibbonshawkingperry} resolve this formally by further contour deformations, and justify this conceptually by interpreting $Z$ not as the true partition function but as an analytic continuation of it from which the density of states can be recovered. But this approach was and is somewhat contentious; see in particular \citeN{grossnucleation} for an alternative interpretation of this term as the instanton controlling nucleation of black holes in `hot flat space'.

In my view the clearest \emph{physical} way to understand what is going on is to follow \citeN{yorkcanonical} in considering a much smaller value of $A_{box}$. (An alternative developed in Brown~\emph{et al}~\citeyear{brownetalblackhole} is to construct a path integral for the \emph{microcanonical} ensemble.) As we have seen, if the box surface lies within the surface of last stable orbit, the canonical ensemble is well-defined and (for large black holes) the thermodynamic limit obtains. Interpreting the path integral in this context and evaluating it to zero order, we get for the Schwarzschild black hole
\be
S_H \equiv \left( \beta \dbd{}{\beta} \right) \ln Z_H(\beta) = \frac{\mc{A}}{4G}
\ee
where $\mc{A}= \beta^2/4\pi$ is the horizon area of a Schwarzschild black hole of mass $M=\beta/8\pi$.

Some observations:
\begin{enumerate}
\item Most obviously and most importantly, this is exactly the classical formula for black hole entropy, for a black hole at the temperature of the ensemble. The statistical-mechanical calculation agrees with the phenomenological results exactly.
\item For black holes much more massive than the Planck length, and temperatures much lower than the Planck temperature, this entropy is much larger than the entropy $S_{flat}$ of flat space. So we are justified in neglecting the $Z_{flat}$ contribution to the entropy, and restricting attention to the black-hole component of the partition function.
\item The area $A_{box}$ of the bounding box does not appear in the expression for entropy. So the microscopic degrees of freedom of the black hole, to zero-loop order, are localised at the horizon. 
\end{enumerate}
I have described these results for the Schwarzschild black hole, but they are quite general. Evaluating the path integral for black holes that are charged and/or rotating again recovers $S_H=\mc{A}/4$. Further, recall (from part I) that \citeN{waldnoether} has used Noether's theorem to construct an expression for the entropy of black holes in generalisations of the Einstein-Hilbert action to arbitrary diffeomorphism-covariant actions. For instance~\cite{demersetal} the entropy of a charged, non-rotating black hole under the generalised action (\ref{generalisedaction}) is (neglecting the higher-order terms not shown)
\be \label{blackholenoether}
S_H = \frac{\mc{A}}{4 G} - 8 \pi u \theta + 16\pi  (1-2u) \gamma
\ee
where $u=r_-/r+$ is the ratio of inner to outer event-horizon radius and parametrises the charge-to-mass ratio of the black hole.
As Wald demonstrates, the zero-loop statistical-mechanical entropy is provably equal to the Noether-charge entropy, despite their completely different conceptual origin: one as  a semiclassical approximation to a count of states in quantum gravity, the other as a theorem in differential geometry.

\subsection{Beyond the zero-loop approximation}\label{oneloop}

Just as in the flat-space case, the one-loop term in the expression (\ref{blackholepartition}) for the black hole partition function has a natural interpretation in terms of the gas of thermal quanta in the black hole's vicinity --- which, for a black hole, is its thermal atmosphere of Hawking-radiation particles. For a sufficiently large box (in the microcanonical ensemble, \iec finessing or ignoring the questions of convergence discussed above) a large part of the entropy of this gas is just the entropy of an ordinary hot gas far from the hole, and is not naturally included in the hole's own entropy; for this reason it is conventional to normalise black hole entropy relative to flat-space entropy, so that we subtract off the entropy of a flat-space gas of quanta in the box. 

On this convention, we nonetheless get a large contribution from the one-loop term. Indeed, naively we get an \emph{infinite} contribution: quanta can be placed arbitrarily close to the horizon, and so formally an arbitrarily large amount of entropy can be stored there. But this is unphysical: according to the effective-field-theory way of understanding QFTs, some cutoff around the Planck length needs to be imposed to regularise the theory. For black holes, this corresponds to assuming that the field-theoretic degrees of freedom exist only outside the stretched horizon (whereas the degrees of freedom identified by the zero-loop calculation live at that horizon). Nonetheless, we might naturally expect large, and cutoff-dependent, additional contributions to the entropy from the one-loop term, potentially conflicting with the value of the entropy deduced from black hole thermodynamics. (Small additional contributions, corresponding to the entropy of ordinary matter a moderate distance from the black hole, might not be problematic; it is the divergent contribution that threatens BHSM.)

This calculation has been explicitly performed (for a free massive scalar field on a charged, unrotating black hole background) by \citeN{demersetal}. Neglecting the small non-divergent terms, they find a one-loop correction equal to
\be
\Delta S=\frac{\mc{A}}{4}\frac{B}{12\pi} + \frac{(2-3u)A}{180}.
\ee
Here $A$ and $B$ are divergently large constants whose exact value depends on the scheme used to cut off the matter-field path integral. In Demers \emph{et al}'s scheme, for instance, 
\be 
A = \ln \left(\frac{4 \mu^2 + m^2}{m^2}\right) + 2 \ln \left(\frac{\mu^2 + m^2}{3\mu^2 + m^2}\right)
\ee
where $m$ is the scalar field mass and $\mu$ is a mass cutoff: $\mu \sim 1/\Lambda$ in the notation of section \ref{qftintro}. (I show the precise form of $A$ more to demonstrate its complexity than anything else.) So the entropy of the black hole, to one-loop order, is
\be
S = \frac{\mc{A}}{4}\left(\frac{1}{G}+\frac{B}{12\pi} \right) - 8\pi u \theta + 16 \pi(1-2u)\gamma  + \frac{(2-3u)A}{180}
\ee
which seems sharply divergent from the phenomenological prediction (\ref{blackholenoether}).

But recall that the empirically-relevant values of the physical constants $G,\theta,\gamma$ are not the bare values that appear in the action, but the renormalised values. To one-loop order, these values are given by
\[
\frac{1}{G_{ren}} = \frac{1}{G}+ \frac{B}{12\pi}
\]
\[
\theta_{ren} = \theta - A/1440\pi \] \be\gamma_{ren} = \gamma + A/1440\pi \ee where $A$, $B$ are the divergently large constants mentioned above. If we re-express the statistical-mechanical value for the entropy in terms of these renormalised constants, we get
\be
S_H = \frac{\mc{A}}{4 G_{ren}} - 8 \pi u \theta_{ren} + 16\pi  (1-2u) \gamma_{ren}.
\ee
which is exactly the Noether-charge entropy expressed in terms of the renormalised coefficients. The only effect of the divergent one-loop terms is to renormalise the constants in the action.

I want to stress the non-triviality of this result (which was originally conjectured by \citeN{susskinduglum} ahead of detailed calculations, and has been reproduced in various other contexts by, \egc, \citeN{solodukhinconical}, \citeN{larsenwilczek}). The form of the renormalisation of $G$, $\theta$ and $\gamma$ follows from considering perturbations around flat spacetime, and involves no input from black hole physics. And yet it is exactly the required value to reproduce the thermodynamic entropy. This is  unmysterious if black hole entropy really does have a statistical-mechanical origin, and if the path-integral approach to quantum gravity is broadly on the right lines. If not, it is inexplicable.

The \emph{non}-divergent part of the one-loop factor can also be evaluated explicitly (see, \egc, \citeN{fursaevconformal,sennonextremal}). It produces corrections to the entropy of form
\be
\Delta S_q = C \ln M
\ee
where $C$ is a dimensionless factor depending on the black hole charge/mass and angular momentum/mass ratios and on the number of massless particles (for the Schwarzschild black hole in the absence of matter, for instance,  $C=199/45$). Since these factors are smaller than the area term in the entropy by a factor $\ln M/M$ --- or about $10^{-36}$ for astrophysical-scale black holes --- they are negligible in the classical limit; as we will see, though, they provide an important precision test of quantum gravity.

\subsection{Black hole pair creation}

Before leaving the low-energy regime, let's consider one more piece of evidence for the statistical-mechanical interpretation of black hole entropy: the quantum-mechanical creation of pairs of charged black holes in a strong electric field. As \citeN{schwingervacuumpolarization} observed, if an electric field is sufficiently strong the particle-free vacuum is unstable against the spontaneous appearance of matter-antimatter pairs of charged particles: the mass-energy cost of creating them is lower than the reduction in the energy of the field due to the charged particles' own fields. Quantitatively, and to zero-loop order (here I follow \citeN[section 5.8]{carlipreview2}) the creation rate for charge-$e$, mass-$m$ particles with $N$ internal states in a field $\vctr{E}$ is
\be
W \sim N\frac{\alpha^2|\vctr{E}|^2}{\pi^2}\exp{- e m^2/|e\vctr{E}|}.
\ee
(This \emph{Schwinger effect} lacks direct experimental confirmation because of the extreme strength of the fields required, but is a prediction of very-well-confirmed physics.) 

If black holes are ordinary statistical-mechanical systems, their rate of creation through the Schwinger process should exceed the production rate of ordinary particles of the same mass and charge by a factor $\exp S_H$, where $S_H$ is the black-hole entropy. This calculation can be done explicitly in the semiclassical approximation within the path-integral approach (the first calculation was by \citeN{garfinklepairproduction}; see \citeN[section 5.8]{carlipreview2} for further references). Exactly the expected result is obtained.

\section{Full quantum gravity}\label{fullquantumgravity}

Low-energy quantum gravity provides a remarkably precise statistical-mechanical reconstruction of black hole thermodynamics, and in doing so provides strong support both for BHSM and for its own validity as a low-energy approximation. Yet it cannot be the whole story, because it is only an \emph{effective} field theory; from here on I consider the more speculative realm of finite, cutoff-free quantum gravity.

\subsection{High-energy and low-energy features of black hole entropy}\label{high-low}

If QMP is correct, then the degrees of freedom counted by black hole entropy are Planckian-scale: they are localised on a stretched horizon $\sim 1$ Planck distance from the true horizon; they are at Planck-scale redshifts compared to infinity; their local temperature is Planckian; there is one degree of freedom per Planck area. So it is perhaps surprising that a low-energy theory of quantum gravity --- effective-field-theory general relativity --- should succeed in giving a statistical-mechanical account of black hole entropy. The surprise can be lessened by remembering that black hole \emph{thermodynamics} was derived in the low-energy regime too, but it is still somewhat mysterious.

The key to the mystery (cf \citeN{susskinduglum}) is to note that black hole entropy is $S=\mc{A}/4G$, where $G$ is the \emph{renormalised} gravitational constant, so that Planck-scale physics shows up in the entropy via the renormalisation process. So low-energy physics constrains the \emph{functional form} of black hole entropy, but its \emph{numerical value} is determined by Planck-scale physics just as are the numerical coefficients of effective field theories. From this perspective, it is perhaps unsurprising that the leading-order term in the path-integral expression for the partition function is a `classical' term with no straightforward interpretation in terms of quantum degrees of freedom (unlike the one-loop correction, which has a fairly transparent quantum interpretation). 

This does not undermine the support that low-energy quantum gravity provides for BHSM: the formula for the partition function is derived microphysically via state counting, after all. But it does mean that any insight into the nature of the microphysical degrees of freedom of a black hole requires a full quantum theory of gravity: that is, a finite theory that reduces to effective field theory in the appropriate regime. So far as I can see it also means that any otherwise-satisfactory theory of full quantum gravity will automatically reproduce the black hole entropy formulae (assuming, that is, that BHSM is correct and the low-energy results are not a bizarre numerical coincidence). So if we possessed multiple theories of full quantum gravity, we could not use their ability to reproduce black hole entropy as a method of selecting between them: failure so to reproduce the entropy would have to mean a more general failure to reproduce low-energy quantum gravity.

At present, though, we have zero theories of full quantum gravity, at least if we require (a) that the theory is at least as cleanly-stated and well-understood as classical general relativity or quantum field theory; (b) that the theory demonstrably reproduces the low-energy limit. What we have are research programs which have partially elucidated hoped-for theories for which we lack a fully explicit statement and understanding.  If we were to calculate black hole entropy (or other features of black-hole thermodynamics) using statistical-mechanical methods within one such research program and got the `wrong' answer, we could reach one of two conclusions: either black hole statistical mechanics is false and there is some non-statistical underpinning for black hole thermodynamics, or else our research program has not succeeded in identifying a viable quantum theory of gravity. Conversely, success in reproducing the entropy would provide (inconclusive) support both for BHSM and for the research program itself. (Not, to be sure, that the quantum theory of gravity promised by that research program is \emph{correct}, but simply that there \emph{is} such a theory and that it is empirically adequate for low-energy physics.)

To keep the scope of this article manageable, I will consider only one quantum theory of gravity: string theory, the locus of the bulk of work both on quantum gravity in general and on black-hole statistical mechanics in particular. (For consideration of BHSM from the point of view of other quantum theories of gravity, see \citeN{carlipreview} and references therein.) String theory is a large and complex subject, and reasons of space (and, frankly, of my own expertise) mean that my account will necessarily be somewhat less detailed and explicit from here on.

\subsection{A lightning review of string theory}

As a starting point for understanding string theory, recall how perturbation theory works for conventional QFT. There, we consider quantum states that can be reasonably-well-approximated as excitations around some classical state, describable by classical fields on a classical spacetime. The excitations are pointlike and can be understood as particles; the bulk (though by no means all) of the phenomenology of QFT can be understood in terms of those excitations and their interactions with one another.

In a sense, `string theory' is less a theory, more a hypothesis about the perturbation theory of quantum gravity. Specifically, string theory makes what we might call the 
\begin{description}
\item[Stringy excitation hypothesis (SEH):] quantum gravity has states that can be approximated as classical (as in QFT) but where the spectrum of elementary excitations around those states include extended (stringy) objects.
\end{description}
A startling amount follows from SEH (here I largely follow \citeN{tongstring}):
\begin{enumerate}
\item Because a string can itself be excited in a great many ways, the spectrum of internal states of the string is very rich, and differently-excited strings can be considered, on lengthscales large compared to the string length, as different species of particles. In particular, one excitation of the string behaves exactly as does the graviton of effective-field-theory general relativity; another describes a scalar particle, the \emph{dilaton}. As such, if SEH holds then pointlike excitations are redundant: \emph{all} elementary excitations can be thought of as extended.
\item The finite size of strings means that they naturally regulate the ultraviolet infinities of point-particle scattering: string-string scattering theory appears to be (though has not been rigorously proved to be) finite to all orders in perturbation theory.
\item Only certain classical states are compatible with the existence of stringy excitations. In particular, the classical spacetime around which a perturbative string theory is built must satisfy the Einstein field equations, or more precisely, must satisfy an equation which is Einstein's equation to first order but has higher-curvature corrections. (And in turn, in the low-curvature regime we can model the classical metric as built up of string excitations, just as a classical electromagnetic field can be built up from quantum excitations of the electromagnetic vacuum.)  More generally, from the interaction structure of the string we can read off the action of the classical solution, and thereby obtain a low-energy effective field theory (which must include at least general relativity) from the string action.
\item Stringy excitations are only possible if the string moves in a quite high number of dimensions: ten, for the supersymmetric strings generally considered in modern string theory. But a lower-dimensional theory can be obtained by (a) supposing that some of those dimensions are compact; (b) considering the low-energy regime where excitations of the string in the compact directions can be neglected. This lets us extract from string theory a wide variety of low-dimensional effective field theories.
\item Between the large number of different ways to compactify higher-dimensional space, and the existence of several different versions of stringy excitations, string theory is capable of underpinning an extremely large number of lower-dimensional field theories, defined on spacetimes of varying dimension and containing varying numbers and kinds of fields. This number is tamed somewhat by the existence of surprising \emph{dualities}, that allow different kinds of excitations around different backgrounds to be regarded as redescriptions of one another, but it remains very large.
\item The average value of the background dilaton field acts as a kind of coupling constant for string-string interactions occurring on that background. As such, the coupling constant can be thought of as varying from solution to solution. In particular, string perturbation theory is expected to be valid in regions where the dilaton field is small; where it is large, string perturbation theory, and perhaps the very description of quantum gravity in terms of stringy excitations, becomes untrustworthy.
\end{enumerate}
Because of string theory's essentially perturbative nature, my statement in the last section (that string theory is a quantum theory of gravity) is not quite correct. It is, rather, a tool which, if SEH is correct, allows us to study some features of quantum gravity in the perturbative regime. To this date, we lack a fully satisfactory theory which is to string theory as QFT is to the perturbative scattering theory of particle physics (though AdS/CFT duality, discussed in section \ref{adscft}, is a partial answer to this problem).

\subsection{Extremal black holes in string theory}\label{string-extremal}

Prima facie, string methods do not seem well suited to black hole statistical mechanics. Perturbative methods work in weakly coupled regimes, and a black hole --- which semiclassically is a region where gravity becomes so strong as to form an event horizon, and quantum-mechanically is a very-strongly-coupled composite system --- scarcely counts as a `weakly-coupled regime'. And in general this is correct: for most black holes, and in particular for astrophysically-relevant examples like the Kerr and Schwarzschild solutions, there is no known way to calculate their statistical-mechanical properties through string-theoretic methods. However, for certain very special cases --- certain sorts of extremal black holes, rather more progress can be made.

I digress briefly to explain the significance of extremal black holes to BHSM. A stationary black hole in four spacetime dimensions, classically, is described by its mass $M$, its angular momentum $J$, and its charge $Q$, but not all values of $M$, $J$ and $Q$ determine a black hole. For sufficiently large values of $J$ and/or $Q$ (for a given mass), the Kerr-Newman solution to the Einstein equations describes not a black hole but a naked singularity: specifically, a black hole is present only if
\be
\frac{J^2}{M^2} + Q^2 \leq M^2.
\ee
 So it is normally assumed that only the region in solution space satisfying this inequality is physical. Points on the edge of this region --- points that exactly saturate the inequality --- are called \emph{extremal black holes}. In higher dimensions, more numbers are required to specify a black hole's charge and angular momentum, but the same basic story applies.
 
 If a charged and/or spinning black hole decays via Hawking radiation, that radiation preferentially includes particles with that sign of charge and angular momentum; however, depending on the details of the particle spectrum, it is often the case that Hawking radiation decreases $M$ more quickly than the other parameters, so that evaporating black holes decay towards extremality. (This is particularly clear for a strongly charged black hole in our universe: there are no massless charged particles in the Standard Model, and massive particles are strongly suppressed in Hawking radiation compared to massless particle, so Hawking radiation tends to increase $|Q/M|$.) And as black holes tend to extremality their surface gravity, and hence their Hawking temperature, approaches zero, so that a black hole will cool arbitrarily close to extremality but not beyond it. Extremal black holes thus play the role of ground states in black hole thermodynamics.
 
 It is a truism of statistical mechanics that systems in their ground state (or very close to it) are much easier to analyse than systems at a finite temperature, and so if we are looking for a statistical-mechanical account of black holes, we might well expect extremal black holes to be a good place to start. And indeed, for at least a large class of extremal black holes in string theory  it is possible to use a rather remarkable trick, first developed by \citeN{stromingervafa}, to calculate their statistical-mechanical properties explicitly.

The details of the trick are well beyond the scope of this paper, but the basic idea is as follows. We can characterise the extremal black hole we are considering by its charges and angular momentum (there may be several charges, as we are potentially considering black holes in the presence of multiple long-range fields). In the weak-coupling regime (where the dilaton field is small) the lowest-energy state with those charges and angular momentum will be some composite of strings which string perturbation theory is well suited to describe, and whose microcanonical entropy (\iec, the dimension of the lowest-energy subspace of Hilbert space) can be calculated fairly straightforwardly. As we turn up the value of the dilaton field, in general we would expect that entropy to change. But in certain special cases it is a topological invariant whose value will remain constant under smooth variation of the dilaton field. Eventually we will have turned it up so much that the system forms a black hole. By now the perturbative description in terms of stringy excitations is hopelessly unreliable --- but the entropy has remained unchanged. The result can then be compared to the thermodynamic entropy; if it does not match, something is badly wrong either with string theory or with BHSM. 

But of course, it does match. Since Strominger and Vafa's original paper, a great deal of work has been done to calculate the statistical-mechanical entropy of extremal black holes, to increasingly high accuracy and in an increasingly large collection of models, via string theory, and in every case the match to the entropy calculated by low-energy methods is perfect. I shall not attempt to summarise this work (see \citeN{mandalsen} and references therein for details) but will give one illustrative example.

The example is a charged, rotating black hole in five spacetime dimensions, obtained by compactifying a certain version of string theory (to be precise, it's type-IIB superstring theory compactified on $K3\times S^1$). This theory is described, in low energies, by a five-dimensional action coupling the metric, a scalar dilaton field, some two-form fields that are analogs of the electromagnetic field, and a variable number $n_v$ of vector fields (the value of $n_v$ depends on how the compactification is carried out). To lowest order the gravitational part of the action is the Einstein-Hilbert action, but there are higher-order terms. Black hole solutions, the analogs of the Kerr-Newman solution, exist for this theory and are parametrised by the mass $M$, the angular momentum $J$, and by three charges that are normally written as $Q_1$, $Q_5$, and $N$. (They are usually called \emph{BMPV} black holes, after \citeN{bmpv}). As with the Kerr-Newman solution, there is an extremal surface in the space of such solutions beyond which naked singularities form; the extremal black holes have nonzero entropy but zero temperature, and can be characterised by $J, Q_1, Q_5, N$, with $M$ given as a function of these.

The area of a BMPV black hole is 
\be
\mc{A}=8 \pi (Q_1Q_5N - J^2)
\ee
and so the entropy to leading order is $\mc{A}/4=2 \pi (Q_1Q_5N - J^2)$. But as we have seen, this entropy receives corrections from two sources: higher-derivative corrections, due to higher-order terms in the gravitational action, and quantum corrections, due to the logarithmic-in-\mc{A} terms in the entropy of the black hole thermal atmosphere. The former can be calculated by Wald's method, the latter by QFT, and the resulting entropy, to highest sub-leading order in each term and for sufficiently large rotating black holes, is
\be
S = 2 \pi (Q_1Q_5N - J^2) (1 + 3/2N) - \frac{1}{12} (n_v-3) \ln (Q_1 Q_5 N - J^2).
\ee
This fairly complicated expression is reproduced, exactly, by string-theoretic calculations: Strominger and Vafa reproduced the leading-order term in the nonrotating limit; Breckenridge~\emph{et al (ibid)} extended it to rotating black holes; \citeN{castromurthy} worked out the higher-derivative terms; \citeN{senextremal} worked out the logarithmic term. 

(As a point of interest, loop quantum gravity can reproduce the leading-order terms in the entropy expression for general black holes (\iec, not just extremal ones) but seems to get the sub-leading quantum corrections \emph{wrong} for at least the Schwarzschild black hole; see \citeN{sennonextremal} for discussion.)

\subsection{Perturbations of extremal black holes}\label{string-nonextremal}

If a (nonrotating) extremal BMPS black hole absorbs a small amount of uncharged mass $\delta M$, it will be perturbed away from equilibrium and acquire a temperature. \citeN{horowitzstrominger} calculated the leading-order change in entropy in this process both via string-theoretic perturbation theory and via the area definition, $\delta S = \delta \mc{A}/4$; they match exactly. 

More dramatically, \citeN{maldacenastrominger} calculated the emission spectrum of a close-to-extremal black hole in string theory, using ordinary statistical-mechanical methods. The result --- grey-body factors and all --- exactly matches the known emission spectrum of that black hole as calculated in ordinary QFT on curved spacetime, even while the form of the calculation is wildly different in the two cases --- the first proceeds via statistical mechanics of stringy excitations, the second by solving the Klein-Gordon equation on a black-hole spacetime. (Apart from supporting BHSM, this provides fairly strong evidence that black hole evaporation is a unitary process; see \citeN[section 4.1]{wallaceinformationloss} for further discussion.)

\section{AdS/CFT duality}\label{adscft}

String theory exactly reproduces the statistical-mechanical description of a large class of extremal and near-extremal black holes, allowing us to derive by statistical-mechanical methods the highly complicated expressions for black hole entropy and evaporation rates that can be derived semiclassically via QFT on curved spacetime, It is pretty difficult to explain this reproduction without accepting that (a) black hole entropy has a statistical-mechanical origin; (b) string theory provides an ultraviolet completion of low-energy quantum general relativity in at least the regimes appropriate to those black holes.

The string-theoretic analysis of extremal black holes was also a major motivation and precursor for the conjectured duality (\citeNP{maldacenaconjecture}, \citeNP{gubseretaladscft}, \citeNP{wittenadscft}) between quantum gravity on anti-de Sitter (AdS) space and conformal quantum field theory (CFT) on the boundary of that space (indeed, in current presentations, the leading-order terms in those calculations are normally reinterpreted in terms of that duality.) But AdS/CFT duality provides evidence for BHSM that goes well beyond the extremal case; in this section I give a brief introduction to AdS/CFT and then consider, in sequence, the evidence for AdS/CFT itself and the evidence it provides for BHSM.

For more details on AdS/CFT see review articles by Aharony \emph{et al}~\citeyear{aharonyadscftreview}, \citeN{harlowreview}, \citeN{hartmanreview}, and \citeN{kaplanadscft} (my account here is based on these sources, especially Harlow and Kaplan). For philosophical discussion, see, \egc, \citeN{deharogaugegravity} or \citeN{tehholography}.

\subsection{Preamble: a note about boundary conditions}

Recall from section \ref{high-low}: in this paper, `low-energy quantum gravity' means general relativity, regarded as an effective field theory and applied only when the energy scales are far below the Planck scale; `full quantum gravity' is any finite theory which reduces to low-energy quantum gravity in appropriately limited regimes. To understand AdS/CFT duality, we also need to consider the \emph{boundary conditions} for low-energy quantum gravity. Consider, in particular, a perturbative regime of low-energy quantum gravity, consisting of perturbations around some classical solution to the field equations on the manifold $\Sigma \times \re$, where the slices $\Sigma \times \{x\}$ are Cauchy surfaces for each $x$. If $\Sigma$ is non-compact, then for the theory to be well-defined we will need to require the perturbations to satisfy certain boundary conditions (normally requiring the perturbations to decay faster than some power of spatial distance; the details will not matter here). This can be reexpressed, and extended beyond the perturbative regime, by interpreting it as a condition on the metrics which are integrated over in the path integral. To say that a given sector of low-energy quantum gravity is (e.g.) defined on `asymptotically Minkowski spacetime' is to say that the path integral satisfies the boundary conditions obtained by considering perturbations around some classical solution that is asymptotically Minkowskian. Note that this is compatible with the path integral being evaluated over solutions of different interior topologies: it is only a constraint on the boundary, not on the interior.

Boundary conditions are also central to understanding the symmetries of a quantum-gravity theory. Recall that classical gravity has the full diffeomorphism group as symmetries, and that this extends to path-integral quantum gravity via the invariance of the action in the path integral. We can divide these diffeomorphisms into three categories:
\begin{enumerate}
\item Those which vanish at the boundary. These correspond to pure gauge redundancy, and act trivially on the Hilbert space of the theory.
\item Those which do not preserve the boundary conditions. These are not represented as transformations at all on the Hilbert space.
\item Those which preserve the boundary conditions but are non-vanishing. These correspond to physical transformations on the system (more precisely: each equivalence class of boundary-preserving diffeomorphisms that differ by a pure-gauge-redundancy diffeomorphism represents a physical transformation) and are represented as nontrivial symmetry transformations on the Hilbert space of the theory. (In the classical limit, they correspond to ADM energy, momentum and the like for asymptotically-Minkowskian boundary conditions.)
\end{enumerate}
In some cosmological contexts --- in particular, for quantum gravity on a compact spatial manifold --- there are no boundary conditions, and all diffeomorphisms are pure gauge redundancies. (The `problem of time' has its sharpest statement in this context, in which it is difficult to write down any gauge-invariant observables at all.) But for black holes, boundary conditions are natural as the hole is normally considered as living in some larger spacetime and the boundary conditions idealise ways in which it can be embedded into such a spacetime; in this context, the physical symmetries can be understood as translating, boosting (etc) the black hole relative to other distant bodies. (See \citeN{wallacerpep} and \citeN{greaveswallacesymmetry} for more discussion on this point.) For AdS/CFT, the relevant boundary condition is for metrics which asymptote to $n$-dimensional \emph{anti-de Sitter} spacetime, $\AdS_n$. This is a maximally homogeneous spacetime which solves the Einstein field equations with negative cosmological constant and no matter. 

The discussion so far has concerned only low-energy quantum gravity, but it is usually assumed (and I will assume it here) that the symmetries of a completion of low-energy quantum gravity include (at least) the symmetries of the low-energy theory identified above. (This is in effect to say that in such theories the diffeomorphism symmetry is not anomalously broken.)

\subsection{Stating the AdS/CFT conjecture}\label{ads-intro}

In its purest and most general form, the AdS/CFT conjecture can be stated as follows:
\begin{quote}
Any full quantum theory of gravity with asymptotically $\AdS_{n+2}$ boundary conditions can be reinterpreted as a conformal quantum field theory on the boundary $S_n\times \re$ of $\AdS_{n+2}$, and vice versa.
\end{quote}
To explain `reinterpreted as' in this statement, it helps to conceive of the quantum gravity theory and the CFT as distinct theories and then see how they are to be identified. In particular, there is a $1:1$ map identifying the Hilbert spaces of the theories, such that under that identification:
\begin{itemize}
\item The global symmetries of the CFT are exactly the asymptotic symmetries of the QG theory, and the generators of symmetry transformations can be identified. (For $n>1$, the symmetry group in both cases is $SO(n,2)$, for $n=1$, it is an infinite-dimensional group.)
\item In particular, the Hamiltonians of the respective theories can be identified, and hence so can their partition functions, and hence all their equilibrium thermodynamic properties.
\item The field operators of the CFT can be defined as (rescaled) limits of field operators of the QG theory as their radial coordinate is taken to infinity. In particular, the limit of the metric tensor of the QG theory is the CFT stress-energy tensor.
\end{itemize}
Over and above this Hilbert-space version of the duality, there is a natural re-expression of the generating function of the CFT  in terms of the partition function of the QG theory.

\subsection{Clarifying the AdS/CFT conjecture}

I now consider three questions each of whose answers (while in each case not entirely uncontentious) hopefully makes clear just how AdS/CFT duality is supposed to function and how it fits into our physics more generally.

\begin{description}
\item[Duality or emergence?] It is natural to think of AdS/CFT correspondence as a relation between two already-understood theories. But we do not possess any non-perturbative, finite description of a QG theory, except via the AdS/CFT correspondence itself. So it is at least a live option to suppose that the right non-perturbative description of a QG theory just is the appropriate CFT, and that there is no independent non-perturbative description. From this perspective we could restate the AdS/CFT conjecture as: any conformal field theory on the boundary of $\AdS_n$ has a low-energy perturbative regime that can be described by low-energy QG on $\AdS_n$, and conversely, any consistent and finite theory that has such a low-energy perturbative description is (interpretable as) a CFT on the boundary of $\AdS_n$. (This statement should make clear that the conjecture remains highly nontrivial on this interpretation.)
\item[Universe or box?] The universe we live in is not AdS spacetime, even approximately (in particular, our universe has positive cosmological constant), and AdS spacetimes are poorly suited for cosmology: they have timelike boundaries, and so require boundary conditions at all times. For this reason, it is generally better to think of the QG theory on asymptotically AdS spacetime not as a model for an entire cosmology, but just as a covariant idealisation for QG in a system confined to a box. The AdS spacetime is then no more physical than the `periodic boundary condition' boxes used in elementary QFT.
\item[Specific or general?] AdS/CFT duality was first discovered \citeN{maldacenaconjecture} in the context of string theory, as a duality between type IIB string theory on $\AdS_5 \times S_5$ and supersymmetric $SU(N)$ gauge theory on the boundary of $\AdS_5$ (call this \emph{Maldacena duality}) and  all other concretely known cases are also string theory/supersymmetric gauge theory dualities. (In each case the `string theory' is characterised only perturbatively and it remains possible that the boundary gauge theory \emph{is} the appropriate non-perturbative description of the string theory.) But AdS/CFT duality in general makes sense for, and can be argued for, in any quantum theory of gravity, and does not rely on particular features of string theory. (See in particular \citeN{kaplanadscft} for a review of AdS/CFT from this perspective.)
\end{description}

\subsection{Why accept AdS/CFT?}

The AdS/CFT conjecture is a conjecture about the mathematical equivalence of two classes of theories, and so `evidence' for it is evidence for a mathematical equivalence, not empirical evidence. Ideally, that `evidence' would consist of a rigorous proof of the duality, but at present we lack a completely rigorous statement of AdS/CFT, let along a proof. There is, however, a great deal of mathematical evidence falling short of proof, which has convinced the great majority of the physics community that the conjecture is correct.  Some illustrative examples include:
\begin{enumerate}
\item Longstanding arguments due to \citeN{thooftgaugestring} suggest strongly that in the large-$N$ limit, $SU(N)$ gauge theory behaves like a string theory; conversely, Maldacena's original argument for AdS/CFT duality proceeds in part through demonstrating explicitly that in a certain limit, the string theory he considers behaves like a conformal gauge field theory. (The `conjecture' is now that these equivalences hold non-perturbatively and away from the limiting cases.)
\item Given a finite field theory on AdS, we can explicitly construct field operators on its boundary from the limits of the AdS fields and verify that they satisfy the conditions for a CFT. If (and only if) the UV-finite field theory is diffeomorphism-invariant and includes a metric field, the constructed CFT will have a conserved stress-energy tensor and thus a well-defined covariant dynamics. 

(By physics standards this might seem to constitute a \emph{proof} of the AdS/CFT conjecture, at least for the AdS $\rightarrow$ CFT direction. The fly in the ointment is that all known quantum field theories of gravity are effective field theories, well-defined only up to some cutoff, and so only partially specify a CFT. The conjecture is then that any short-distance completion of the QG theory will be reinterpretable as a completion of the partially-specified CFT.)
\item Conversely, we can give explicit CFT descriptions of at least some AdS quantum states: specifically, those corresponding to multiparticle perturbations of the AdS vacuum where the particle density is low compared to the Planck density. These constructions are valid only perturbatively and it remains obscure how to give a CFT specification of strongly coupled states of the interior.
\item Much of the power of the AdS/CFT correspondence is that strong-field regimes on the AdS side correspond to weak-field regimes on the CFT side and vice versa, which means that calculations that are easy on one side are usually intractable on the other. But there are nonetheless many examples of quantities which \emph{can} be calculated explicitly both on the AdS and the CFT sides of the duality and which match exactly --- and, just as importantly, no known cases where these calculations give inconsistent answers. In one particularly striking example from Maldacena duality (which I borrow from \citeN[pp.120-122]{conlonwhystringtheory}), a certain quantity --- the anomalous dimension of the `Konishi operator' can be calculated via perturbation theory on both sides of the duality. 

The calculation has been done to five-loop order both in string theory on $\AdS_5 \times S_5$ and in the associated conformal field theory in the large-$N$ limit (Bajnok~\emph{et al}~\citeyearNP{bajnokkonishi}, \citeNP{balogkonishi}, Eden~\emph{et al}~\citeyearNP{edenkonishi}). The calculations look completely different but the result, in either case, is
\[
\gamma_K(a) = 3a - 3a^2 +\frac{21}{4}a^3 - \left( \frac{39}{4} - \frac{9}{4}\zeta(3) + \frac{45}{8}\zeta(5)\right)a^4\]
\be
+  \left(\frac{237}{16}- \frac{81 }{16}\zeta(3)^2 + \frac{27}{4} \zeta(3) - \frac{135}{16}  \zeta(5)+ \frac{945}{32} \zeta(7)  \right)a^5
\ee
where $\zeta$ is the Riemann zeta function and $a$ is the appropriate coupling constant. It is hard, to put it mildly, to believe that it is just a coincidence that this complicated expression is obtained by two completely different calculational methods in the absence of a duality between the underlying theories. (I invite a sceptic to discuss terms for a bet on whether the \emph{six-loop} calculations, as and when they are done, continue to match!)
\end{enumerate}
Reasons like these seem to give pretty good grounds to provisionally accep the AdS/CFT conjecture and ask what it tells us about black hole statistical mechanics and black hole evaporation. And indeed --- as we shall see --- the impressive consistency of the story that it tells is itself further reason to take the correspondence seriously.

\subsection{Phase transitions in bulk and boundary}

Black holes can exist in AdS space just as in Minkowski space (more carefully: black holes can have asymptotically AdS boundary conditions as well as asymptotically-Minkowski boundary conditions). When their Schwarzschild radius is small compared to the effective radius of the space, they are pretty much identical to asymptotically Minkowski black holes, though they differ somewhat at larger sizes. The standard arguments for black hole thermodynamics (the area theorem, reversible interactions, Hawking radiation, etc) go across \emph{mutatis mutandis} and recover the same formula for entropy as in the Minkowski case: to leading order and in $G=1$ units, the entropy is $1/4$ of the horizon area, with corrections due to higher-order terms in the gravitational Lagrangian.

I mentioned in section \ref{ads-intro} that AdS space functions as a box, and indeed the thermodynamics of an AdS black hole is qualitatively the same as for a Minkowski black hole in a box. Small black holes are thermodynamically unstable and will radiate away; sufficiently large black holes are thermodynamically stable, both if the space is taken as isolated and if it is coupled to an external heat reservoir (though the criterion for `sufficiently large' is more demanding in the latter case). There is therefore a phase transition --- the \emph{Hawking-Page transition} \cite{hawkingpage} --- as the energy of the box is increased: at low energies, the equilibrium state is a gas of gravitons and matter particles (and strings, if the stringy excitation hypothesis is correct); at high energies, it is a single large black hole. And the functional dependence of entropy on energy has a different, known, form in each phase. At high energies, in particular, the entropy of an $n$-dimensional black hole scales as $M^{d-2}$. 

If black hole thermodynamics does have a statistical-mechanical underpinning, then (since the statistical mechanics of bulk and boundary theories are identical) we should expect to see this phase transition, and this functional form of the entropy, in the CFT description. And indeed we do. The CFTs that (seem to) correspond to realistic quantum theories of gravity are non-Abelian gauge theories, and interpreted that way, the Hawking-Page transition is a \emph{confinement-deconfinement} transition. At high energies, non-Abelian gauge theories can be approximated as free-field theories, and so the high-energy regime can be treated as a radiation gas, with $S\propto M^{d-2}$ just as in the bulk description. At low energies, color confinement shifts the gauge theory into a regime of strongly coupled color singlets. In general, the complexity of the theory in this strong-coupling regime makes it impossible to calculate the quantitative details of this phase transition, but the qualititative picture is fairly well understood (see in particular Aharony \emph{et al}~\citeyear{hagedorndeconfinementadscft}) and matches the bulk description very accurately.

\subsection{Quantitative results from $\AdS_3/\CFT_2$}

In some cases, we can do better than simply reproducing the qualitative features of black hole thermodynamics. Quantum gravity on $\AdS_3$ is dual to a two-dimensional conformal field theory, and the latter is fairly well understood (partly because the conformal group in two dimensions is much larger than in three or more dimensions and so constrains the theory more sharply). In particular, 2-dimensional CFTs can be characterised by their \emph{central charge} $c$ (a parameter which \emph{inter alia} parametrises the way in which the representation of the 2D conformal group in the CFT is a projective representation) and the celebrated \emph{Cardy formula} \cite{cardyformula} gives the partition function of a general 2D CFT on a circle of unit radius, in the high-temperature limit, as a function of $c$:
\be
Z(\beta)=\e{\pi^2 c/3\beta}.
\ee
(See \citeN[section 6.4]{harlowreview} for more details.)
Since the central charge just depends on the theory's symmetry structure, it can be calculated on the AdS side; if the result is plugged into the Cardy formula and the entropy calculated, the partition function of the CFT is precisely what the partition function of the quantum-gravity theory would have to be if the thermodynamic description of the black hole had a statistical-mechanical basis; in particular, the Bekenstein-Hawking entropy formula is recovered exactly. This would again be an inexplicable coincidence unless (a) there really is a duality between quantum gravity on $\AdS_3$ and some CFT on its boundary, and (b) black holes really are statistical-mechanical systems. (Furthermore, the temperature below which the Cardy formula fails matches the temperature of the Hawking/Page phase transition, although it cannot be calculated precisely with present methods.)

These results can be extended. A large class of extremal black holes (even in asymptotically-Minkowski space) can be approximated, close to their horizon, by $\AdS\times K$ for some compact $K$. Applying the correspondence to this region, we can again recover the central charge, plug it into the Cardy formula, and compare the result to the black hole's thermodynamic entropy; in each case, the match is exact. (Indeed, the Strominger-Vafa calculations in string theory can be reinterpreted via the AdS/CFT correspondence as proceeding in this way, as observed by \citeN{stromingerconformal}.) This description even gives some (rather heuristic) insight as to the nature of the microscopic degrees of freedom counted by black hole entropy: they live in a conformal field theory at the `boundary' of the near-horizon region, \iec on the stretched horizon, in support of QMP. Other aspects of sections \ref{string-extremal}--\ref{string-nonextremal}'s discussion can also be reinterpreted via AdS/CFT, in particular the Maldacena-Strominger results on decay rates of near-extremal black holes; see \citeN[sections 12-13]{hartmanreview} for further discussion.

The `large class of extremal black holes' discussed above does not include physically realistic black holes in our universe. However, the \emph{Kerr/CFT correspondence} (Guica~\emph{et al}~\citeyearNP{guicaetal}) appears to extend the basic idea (a duality between the near-horizon region of an extremal black hole and a conformal field theory on the boundary of that region) to extremal Kerr black holes in four dimensions. The mathematical and physical basis of this extended  correspondence is distinctly less clear than conventional AdS/CFT, but calculations have again achieved precise quantitative agreement with the thermodynamic properties of these black holes as calculated semiclassically (cf Bredberg~\emph{et al}~\citeyear{bredbergetal} and references therein). Since there are astrophysical black holes that can be approximated pretty well as extremal Kerr black holes, this is tantalisingly close to reproducing the statistical mechanics of directly-observed black holes. Furthermore, there is calculational evidence that the Kerr/CFT correspondence extends to \emph{non}-extremal Kerr black holes --- that is, to basically all astrophysical black holes! --- although the conceptual basis of this extension is far from clear. The Kerr/CFT correspondence is cutting-edge theoretical physics and it would be premature to review it further in an article of this kind, but it is clearly an exciting development. (For an introduction, see Bredberg \emph{et al}, \emph{ibid}; for a detailed and up-to-date review, see \citeN{comperekerrcft}.)

\section{Conclusions}\label{conclusion}

\begin{quote}
In practice, the hard acid that dissolves scepticism is the ability to calculate.

\begin{flushright}
Joseph Conlon, \emph{Why String Theory?}\footnote{\citeN[p.6]{conlonwhystringtheory}}
\end{flushright}
\end{quote}
Black hole entropy can be calculated:
\begin{enumerate}
\item In differential geometry, through the First Law of Black Hole thermodynamics (in Wald's extended form);
\item To one-loop order in low-energy quantum gravity, through the path integral formalism;
\item In string theory, for a large class of extremal black holes;
\item Via the AdS/CFT correspondence, qualitatively in the general case and quantitatively in a number of cases.
\end{enumerate}
Whenever the calculation is done by two or more methods, exact agreement is obtained, including in many cases not merely the functional form (`entropy is proportional to area') or the leading-order result (`entropy=area/4'), but the complicated, detailed subleading-order corrections to the area formula.

It is logically possible that this is all a trick of the light. Nothing I have discussed here makes it \emph{impossible} that low-energy quantum gravity is hopelessly ill-defined or has nothing to do with semiclassical physics; that no theory of quantum gravity satisfies the stringy excitation hypothesis; that the AdS/CFT conjecture is false; that black hole thermodynamics has no statistical-mechanical underpinning. 

But `logically possible' is not the same as `remotely plausible'. It is not remotely plausible that all of this is a massive coincidence. And while it cannot be ruled out logically that there is some baroque alternative explanation, overwhelmingly the most natural explanation is: black hole thermodynamics is underpinned by statistical mechanics just like any other thermodynamics; general relativity can be analysed by QFT at low energies just like any other field theory, and its quantum statistical mechanics thus reproduces black hole entropy; at least some consistent high-energy completion of effective-field-theory gravity satisfies the stringy excitation hypothesis, and so string theory likewise reproduces the statistical-mechanical results; the AdS/CFT conjecture is true, so that black hole statistical mechanics is dual to the statistical mechanics of a conformal field theory in at least some important cases.

Philosophers of science distinguish between the context of discovery where theories are invented, and the context of justification where they are tested. Quantum gravity remains firmly in the context of discovery: we are a long way from a cleanly-formulated theory which reproduces all the predictions of sub-Planck-level physics, and even further from one which makes novel and testable predictions that transcend the low-energy regime. So any arguments in this field are necessarily somewhat tentative. But as tentative hypotheses go, the hypothesis that black holes are statistical-mechanical systems seems as solid as any other result we have in quantum gravity, and more solid than most. 

And this leads to a notorious paradox. If black holes are ordinary statistical-mechanical systems obeying unitary quantum mechanics, that strongly suggests that their radiation and eventual evaporation likewise has a fully unitary description. But the calculations of Hawking radiation in QFT --- the very calculations which got the field of black hole thermodynamics and statistical mechanics off the ground --- describe a radiative process that appears strongly non-unitary. I discuss the conceptual structue this \emph{information loss paradox}, within the framework for BHSM developed above, in \citeN{wallaceinformationloss}.


\begin{thebibliography}{}

\bibitem[\protect\citeauthoryear{Aharony, Gubser, Maldacena, Ooguri, and
  Oz}{Aharony et~al.}{1999}]{aharonyadscftreview}
Aharony, O., S.~Gubser, J.~Maldacena, H.~Ooguri, and Y.~Oz (1999).
\newblock Large {N} field theories, string theory and gravity.
\newblock {\em Physics Reports\/}~{\em 323}, 183--286.

\bibitem[\protect\citeauthoryear{Aharony, Marsano, Minwalla, Papadodimas, and
  Van~Raamsdonk}{Aharony et~al.}{2004}]{hagedorndeconfinementadscft}
Aharony, O., J.~Marsano, S.~Minwalla, K.~Papadodimas, and M.~Van~Raamsdonk
  (2004).
\newblock The {H}agedorn/deconfinement phase transition in weakly coupled large
  {N} gauge theories.
\newblock {\em Advances in Theoretical and Mathematical Physics\/}~{\em 8},
  603--696.

\bibitem[\protect\citeauthoryear{Bajnok, Hegedus, Janik, and Lukowski}{Bajnok
  et~al.}{2010}]{bajnokkonishi}
Bajnok, Z., A.~Hegedus, R.~A. Janik, and T.~Lukowski (2010).
\newblock Five loop {K}onishi from {A}d{S}/{CFT}.
\newblock {\em Nuclear Physics B\/}~{\em 827}, 426--456.

\bibitem[\protect\citeauthoryear{Balog and Hegedus}{Balog and
  Hegedus}{2010}]{balogkonishi}
Balog, J. and A.~Hegedus (2010).
\newblock 5-loop {K}onishi from linearized {TBA} and the {XXX} magnet.
\newblock {\em JHEP\/}~{\em 1006}, 080.

\bibitem[\protect\citeauthoryear{Banks}{Banks}{2008}]{banksQFT}
Banks, T. (2008).
\newblock {\em Modern Quantum Field Theory: a Concise Introduction}.
\newblock Cambridge: Cambridge University Press.

\bibitem[\protect\citeauthoryear{Breckenridge, Myers, Peet, and
  Vafa}{Breckenridge et~al.}{1997}]{bmpv}
Breckenridge, J., R.~Myers, A.~Peet, and C.~Vafa (1997).
\newblock D-branes and spinning black holes.
\newblock {\em Physics Letters B\/}~{\em 391}, 93--98.

\bibitem[\protect\citeauthoryear{Bredberg, Keeler, Lysov, and
  Strominger}{Bredberg et~al.}{2011}]{bredbergetal}
Bredberg, I., C.~Keeler, V.~Lysov, and A.~Strominger (2011).
\newblock {C}argese lectures on the {K}err/{CFT} correspondence.
\newblock {\em Nuclear Physics Proceedings Supplement\/}~{\em 216}, 194--210.

\bibitem[\protect\citeauthoryear{Brown, Comer, Martinez, Melmed, Whiting, and
  York~Jr}{Brown et~al.}{1990}]{brownetalblackhole}
Brown, J., G.~Comer, E.~Martinez, J.~Melmed, B.~Whiting, and J.~York~Jr (1990).
\newblock Thermodynamic ensembles and gravitation.
\newblock {\em Classical and Quantum Gravity\/}~{\em 7}, 1433--1444.

\bibitem[\protect\citeauthoryear{Burgess}{Burgess}{2004}]{burgesseffectivegr}
Burgess, C. (2004).
\newblock Quantum gravity in everyday life: General relativity as an effective
  field theory.
\newblock {\em Living Reviews in Relativity\/}~{\em 7}, 5.

\bibitem[\protect\citeauthoryear{Cardy}{Cardy}{1986}]{cardyformula}
Cardy, J. (1986).
\newblock Operator content of two-dimensional conformally invariant theories.
\newblock {\em Nuclear Physics B\/}~{\em 270}, 186--204.

\bibitem[\protect\citeauthoryear{Carlip}{Carlip}{2008}]{carlipreview}
Carlip, S. (2008).
\newblock Black hole thermodynamics and statistical mechanics.
\newblock https://arxiv.org/abs/0807.4520.

\bibitem[\protect\citeauthoryear{Carlip}{Carlip}{2014}]{carlipreview2}
Carlip, S. (2014).
\newblock Black hole thermodynamics.
\newblock http://arxiv.org/abs/1410.1486.

\bibitem[\protect\citeauthoryear{Castro and Murthy}{Castro and
  Murthy}{2009}]{castromurthy}
Castro, A. and S.~Murthy (2009).
\newblock Corrections to the statistical entropy of five dimensional black
  holes.
\newblock {\em JHEP\/}~{\em 0906}, 024.

\bibitem[\protect\citeauthoryear{Compere}{Compere}{2017}]{comperekerrcft}
Compere, G. (2017).
\newblock The {K}err/{CFT} correspondence and its extensions.
\newblock {\em Living Reviews in Relativity\/}~{\em 20}, 1.

\bibitem[\protect\citeauthoryear{Conlon}{Conlon}{2016}]{conlonwhystringtheory}
Conlon, J. (2016).
\newblock {\em Why String Theory?}
\newblock London: Taylor and Francis.

\bibitem[\protect\citeauthoryear{De~Haro, Mayerson, and Butterfield}{De~Haro
  et~al.}{2016}]{deharogaugegravity}
De~Haro, S., D.~R. Mayerson, and J.~N. Butterfield (2016).
\newblock Conceptual aspects of gauge/gravity duality.
\newblock https://arxiv.org/abs/1509.09231.

\bibitem[\protect\citeauthoryear{Demers, Lafrance, and Myers}{Demers
  et~al.}{1995}]{demersetal}
Demers, J.-G., R.~Lafrance, and R.~Myers (1995).
\newblock Black hole entropy without brick walls.
\newblock {\em Physical Review D\/}~{\em 52}, 2245--2253.

\bibitem[\protect\citeauthoryear{Duncan}{Duncan}{2012}]{duncanQFT}
Duncan, T. (2012).
\newblock {\em The Conceptual Framework of Quantum Field Theory}.
\newblock Oxford: Oxford University Press.

\bibitem[\protect\citeauthoryear{Eden, Heslop, Korchemsky, Smirnov, and
  Sokatchev}{Eden et~al.}{2012}]{edenkonishi}
Eden, B., P.~Heslop, G.~P. Korchemsky, V.~A. Smirnov, and E.~Sokatchev (2012).
\newblock Five-loop {K}onishi in {N}=4 {SYM}.
\newblock https://arxiv.org/abs/1202.5733.

\bibitem[\protect\citeauthoryear{Fursaev}{Fursaev}{1995}]{fursaevconformal}
Fursaev, D. (1995).
\newblock Temperature and entropy of a quantum black hole and conformal
  anomaly.
\newblock {\em Physical Review D\/}~{\em 51}, 5352.

\bibitem[\protect\citeauthoryear{Garfinkle, Giddings, and Strominger}{Garfinkle
  et~al.}{1994}]{garfinklepairproduction}
Garfinkle, D., S.~Giddings, and A.~Strominger (1994).
\newblock Entropy in black hole pair production.
\newblock {\em Physical Review D\/}~{\em 49}, 958--965.

\bibitem[\protect\citeauthoryear{Gibbons and Hawking}{Gibbons and
  Hawking}{1977}]{gibbonshawking}
Gibbons, G. and S.~Hawking (1977).
\newblock Action integrals and partition functions in quantum gravity.
\newblock {\em Physical Review D\/}~{\em 15}, 2752--2756.

\bibitem[\protect\citeauthoryear{Gibbons, Hawking, and Perry}{Gibbons
  et~al.}{1978}]{gibbonshawkingperry}
Gibbons, G., S.~Hawking, and M.~Perry (1978).
\newblock Path integrals and the indefiniteness of the gravitational action.
\newblock {\em Nuclear Physics B\/}~{\em 138}, 141--150.

\bibitem[\protect\citeauthoryear{Greaves and Wallace}{Greaves and
  Wallace}{2014}]{greaveswallacesymmetry}
Greaves, H. and D.~Wallace (2014).
\newblock Empirical consequences of symmetries: a new framework.
\newblock {\em British Journal for the Philosophy of Science\/}~{\em 65},
  59--89.

\bibitem[\protect\citeauthoryear{Gross, Perry, and Yaffe}{Gross
  et~al.}{1982}]{grossnucleation}
Gross, D.~J., M.~J. Perry, and L.~G. Yaffe (1982).
\newblock Instability of flat space at finite temperature.
\newblock {\em Physical Review D\/}~{\em 25}, 330--355.

\bibitem[\protect\citeauthoryear{Gubser, Klebanov, and Polyakov}{Gubser
  et~al.}{1998}]{gubseretaladscft}
Gubser, S., I.~Klebanov, and A.~Polyakov (1998).
\newblock Gauge theory correlators from non-critical string theory.
\newblock {\em Physics Letters\/}~{\em B428}, 105--114.

\bibitem[\protect\citeauthoryear{Guica, Hartman, Song, and Strominger}{Guica
  et~al.}{2009}]{guicaetal}
Guica, M., T.~Hartman, W.~Song, and A.~Strominger (2009).
\newblock The {K}err/{CFT} correspondence.
\newblock {\em Physical Review D\/}~{\em 80}, 124008.

\bibitem[\protect\citeauthoryear{Harlow}{Harlow}{2016}]{harlowreview}
Harlow, D. (2016).
\newblock Jerusalem lectures on black holes and quantum information.
\newblock {\em Reviews of Modern Physics\/}~{\em 88}, 015002.

\bibitem[\protect\citeauthoryear{Hartle and Schleich}{Hartle and
  Schleich}{1987}]{hartleschleich}
Hartle, J. and K.~Schleich (1987).
\newblock The conformal rotation in linearised gravity.
\newblock In I.~Batalin, C.~Isham, and G.~Vikovisky (Eds.), {\em Quantum field
  theory and quantum statistics: essays in honour of the sixtieth birthday of
  {E}. {S}. {F}radkin}, Volume~2, pp.\  67--87. Bristol: Adam Hilger.

\bibitem[\protect\citeauthoryear{Hartman}{Hartman}{2015}]{hartmanreview}
Hartman, T. (2015).
\newblock Lectures on quantum gravity and black holes.
\newblock http://www.hartmanhep.net/topics2015/gravity-lectures.pdf.

\bibitem[\protect\citeauthoryear{Hawking}{Hawking}{1975}]{hawking1975}
Hawking, S. (1975).
\newblock Particle creation by black holes.
\newblock {\em Communications in Mathematical Physics\/}~{\em 43}, 199.

\bibitem[\protect\citeauthoryear{Hawking and Page}{Hawking and
  Page}{1983}]{hawkingpage}
Hawking, S. and D.~N. Page (1983).
\newblock Thermodynamics of black holes in anti-de {S}itter space.
\newblock {\em Communications in Mathematical Physics\/}~{\em 87}, 577--588.

\bibitem[\protect\citeauthoryear{Horowitz and Strominger}{Horowitz and
  Strominger}{1996}]{horowitzstrominger}
Horowitz, G. and A.~Strominger (1996).
\newblock Counting states of near-extremal black holes.
\newblock {\em Physical Review Letters\/}~{\em 77}, 2368--2371.

\bibitem[\protect\citeauthoryear{Kaplan}{Kaplan}{2016}]{kaplanadscft}
Kaplan, J. (2016).
\newblock Lectures on {A}d{S}/{CFT} from the bottom up.
\newblock
  http://sites.krieger.jhu.edu/jared-kaplan/files/2016/05/AdSCFTCourseNotesCurrentPublic.pdf.

\bibitem[\protect\citeauthoryear{Larsen and Wilczek}{Larsen and
  Wilczek}{1996}]{larsenwilczek}
Larsen, F. and F.~Wilczek (1996).
\newblock Renormalization of black hole entropy and of the gravitational
  coupling constant.
\newblock {\em Nuclear Physics B\/}~{\em 458}, 249--266.

\bibitem[\protect\citeauthoryear{Maldacena and Strominger}{Maldacena and
  Strominger}{1997}]{maldacenastrominger}
Maldacena, J. and A.~Strominger (1997).
\newblock Black hole greybody factors and {D}-brane spectroscopy.
\newblock {\em Physical Review D\/}~{\em 55}, 861--870.

\bibitem[\protect\citeauthoryear{Maldacena}{Maldacena}{1998}]{maldacenaconjecture}
Maldacena, J.~M. (1998).
\newblock The large {N} limit of superconformal field theories and
  supergravity.
\newblock {\em Advances in Theoretical and Mathematical Physics\/}~{\em 2},
  231--232.

\bibitem[\protect\citeauthoryear{Mandal and Sen}{Mandal and
  Sen}{2010}]{mandalsen}
Mandal, I. and A.~Sen (2010).
\newblock Black hole microstate counting and its macroscopic counterpart.
\newblock https://arxiv.org/abs/1008.3801.

\bibitem[\protect\citeauthoryear{Polchinski}{Polchinski}{2016}]{polchinskiblackholereview}
Polchinski, J. (2016).
\newblock The black hole information problem.
\newblock https://arxiv.org/abs/1609.04036.

\bibitem[\protect\citeauthoryear{Reuter and Saueressig}{Reuter and
  Saueressig}{2012}]{reutersaueressig}
Reuter, M. and F.~Saueressig (2012).
\newblock Quantum {E}instein gravity.
\newblock https://arxiv.org/abs/1202.2274.

\bibitem[\protect\citeauthoryear{Schleich}{Schleich}{1987}]{schleich}
Schleich, K. (1987).
\newblock Conformal rotation in perturbative gravity.
\newblock {\em Physical Review D\/}~{\em 36}, 2342.

\bibitem[\protect\citeauthoryear{Schwinger}{Schwinger}{1951}]{schwingervacuumpolarization}
Schwinger, J. (1951).
\newblock On gauge invariance and vacuum polarization.
\newblock {\em Physical Review\/}~{\em 82}, 664--679.

\bibitem[\protect\citeauthoryear{Sen}{Sen}{2011}]{senextremal}
Sen, A. (2011).
\newblock Logarithmic corrections to rotating extremal black hole entropy in
  four and five dimensions.
\newblock https://arxiv.org/abs/1109.3706.

\bibitem[\protect\citeauthoryear{Sen}{Sen}{2013}]{sennonextremal}
Sen, A. (2013).
\newblock Logarithmic corrections to {S}chwarzschild and other non-extremal
  black hole entropy in different dimensions.
\newblock {\em JHEP\/}~{\em 04}, 156.

\bibitem[\protect\citeauthoryear{Solodukhin}{Solodukhin}{1995}]{solodukhinconical}
Solodukhin, S. (1995).
\newblock The conical singularity and quantum corrections to entropy of black
  hole.
\newblock {\em Physical Review D\/}~{\em 51}, 609--617.

\bibitem[\protect\citeauthoryear{Srednicki}{Srednicki}{2007}]{srednickiqft}
Srednicki, M. (2007).
\newblock {\em Quantum Field Theory}.
\newblock Cambridge: Cambridge University Press.

\bibitem[\protect\citeauthoryear{Strominger}{Strominger}{1998}]{stromingerconformal}
Strominger, A. (1998).
\newblock Black hole entropy from near-horizon microstates.
\newblock {\em JHEP\/}~{\em 9802}, 009.

\bibitem[\protect\citeauthoryear{Strominger and Vafa}{Strominger and
  Vafa}{1996}]{stromingervafa}
Strominger, A. and C.~Vafa (1996).
\newblock Microscopic origin of the {B}ekenstein-{H}awking entropy.
\newblock {\em Physics Letters\/}~{\em B379}, 99--104.

\bibitem[\protect\citeauthoryear{Susskind, Thorlacius, and Uglum}{Susskind
  et~al.}{1993}]{susskindthorlaciusuglum}
Susskind, L., L.~Thorlacius, and J.~Uglum (1993).
\newblock The stretched horizon and black hole complementarity.
\newblock {\em Physical Review D\/}~{\em 48}, 3743--3761.

\bibitem[\protect\citeauthoryear{Susskind and Uglum}{Susskind and
  Uglum}{1994}]{susskinduglum}
Susskind, L. and J.~Uglum (1994).
\newblock Black hole entropy in canonical quantum gravity and superstring
  theory.
\newblock {\em Physical Review D\/}~{\em 50}, 2700--2711.

\bibitem[\protect\citeauthoryear{'t~Hooft}{'t~Hooft}{1974}]{thooftgaugestring}
't~Hooft, G. (1974).
\newblock A planar diagram theory for strong interactions.
\newblock {\em Nuclear Physics B\/}~{\em 72}, 461.

\bibitem[\protect\citeauthoryear{'t~Hooft}{'t~Hooft}{1985}]{thooftblackhole}
't~Hooft, G. (1985).
\newblock On the quantum structure of a black hole.
\newblock {\em Nuclear Physics B\/}~{\em 256}, 727--745.

\bibitem[\protect\citeauthoryear{Teh}{Teh}{2013}]{tehholography}
Teh, N.~J. (2013).
\newblock Holography and emergence.
\newblock {\em Studies in History and Philosophy of Modern Physics\/}~{\em 44},
  300--311.

\bibitem[\protect\citeauthoryear{Thorne, Price, and Macdonald}{Thorne
  et~al.}{1986}]{membraneparadigm}
Thorne, K.~S., R.~H. Price, and D.~A. Macdonald (Eds.) (1986).
\newblock {\em Black Holes: the Membrane Paradigm}.
\newblock New Haven: Yale University Press.

\bibitem[\protect\citeauthoryear{Tong}{Tong}{2009}]{tongstring}
Tong, D. (2009).
\newblock Lectures on string theory.
\newblock https://arxiv.org/abs/0908.0333.

\bibitem[\protect\citeauthoryear{Wald}{Wald}{1993}]{waldnoether}
Wald, R. (1993).
\newblock Black hole entropy is {N}oether charge.
\newblock {\em Physical Review D\/}~{\em 48}, 3427--3431.

\bibitem[\protect\citeauthoryear{Wallace}{Wallace}{2017a}]{wallaceblackholethermodynamics}
Wallace, D. (2017a).
\newblock The case for black hole thermodynamics, part {I}: phenomenological
  thermodynamics.
\newblock Forthcoming in \emph{Studies in the History and Philosophy of Modern
  Physics.} https://arxiv.org/abs/1710.02724.

\bibitem[\protect\citeauthoryear{Wallace}{Wallace}{2017b}]{wallaceqfthandbook}
Wallace, D. (2017b).
\newblock Quantum field theory.
\newblock Forthcoming in Knox and Wilson (ed.), \emph{Handbook of the
  Philosophy of Physics} (Routledge).

\bibitem[\protect\citeauthoryear{Wallace}{Wallace}{2017c}]{wallacerpep}
Wallace, D. (2017c).
\newblock The relativity and equivalence principles for self-gravitating
  systems.
\newblock In D.~Lehmkuhl, G.~Schliemann, and E.~Scholz (Eds.), {\em Towards a
  theory of spacetime theories}, pp.\  257--266. New York: Birkh{\"a}user.

\bibitem[\protect\citeauthoryear{Wallace}{Wallace}{2017d}]{wallaceinformationloss}
Wallace, D. (2017d).
\newblock Why information loss is paradoxical.
\newblock To appear in N. Huggett, K. Matsubara and C. Wuthrich (eds.)
  \emph{Beyond Spacetime} (Cambridge University Press, forthcoming).
  https://arxiv.org/abs/1710.03783.

\bibitem[\protect\citeauthoryear{Weinberg}{Weinberg}{1979}]{weinbergasymptotic}
Weinberg, S. (1979).
\newblock Ultraviolet divergences in quantum theories of gravitation.
\newblock In S.~Hawking and W.~Israel (Eds.), {\em General Relativity: an
  {E}instein Centenary Survey}, pp.\  790--832. Cambridge: Cambridge University
  Press.

\bibitem[\protect\citeauthoryear{Weinberg}{Weinberg}{2008}]{weinbergcosmology}
Weinberg, S. (2008).
\newblock {\em Cosmology}.
\newblock Oxford: Oxford University Press.

\bibitem[\protect\citeauthoryear{Witten}{Witten}{1998}]{wittenadscft}
Witten, E. (1998).
\newblock Anti de {S}itter space and holography.
\newblock {\em Advances in Theoretical and Mathematical Physics\/}~{\em 2},
  253--291.

\bibitem[\protect\citeauthoryear{York}{York}{1986}]{yorkcanonical}
York, J.~W. (1986).
\newblock Black-hole thermodynamics and the {E}uclidean {E}instein action.
\newblock {\em Physical Review D\/}~{\em 33}, 2092--2099.

\bibitem[\protect\citeauthoryear{Zee}{Zee}{2003}]{zeeQFT}
Zee, A. (2003).
\newblock {\em Quantum Field Theory in a Nutshell}.
\newblock Princeton: Princeton University Press.

\end{thebibliography}

\end{document}